\documentclass[9pt,twocolumn,twoside]{pnas-new-arxive}

\usepackage[T1]{fontenc}
\usepackage[symbol*]{footmisc}
\setfnsymbol{bringhurst}

\templatetype{pnasresearcharticle} 

\usepackage{comment}
\usepackage{xspace}
\newcommand{\eq}[1]{eq.~(\ref{#1})}

\newcommand{\fig}[1]{Fig.~\ref{#1}}
\newcommand{\quot}[1]{``#1''}

\newcommand{\vs}{\textrm{vs.}}

\newcommand{\OCAL}{\mathcal{O}}  
\newcommand{\expc}[1]{\exp \glc #1 \grc} 
\newcommand{\glb}{\left(}  
\newcommand{\grb}{\right)}  
\newcommand{\glc}{\left[}  
\newcommand{\grc}{\right]}  

\newcommand{\ahat}{\hat{a}}
\newcommand{\bhat}{\hat{b}}

\newcommand{\xhat}{\hat{x}}

\newcommand{\mean}[1]{\left\langle #1 \right\rangle}

\newcommand\bigOb[1]{\ensuremath{\OCAL\glb #1 \grb}}

\newcommand{\fraca}[2]{#1 / #2}
\newcommand{\fracb}[2]{\frac{#1}{#2}}

\newcommand\subfig[2]{{Fig.~\ref{#1}{#2}}}
\newcommand\subcap[1]{{(#1):}}

\newcommand{\Ulj}{U_{\text{LJ}}}
\newcommand{\lch}{l_c}

\newcommand{\REF}[2][]{\ifthenelse{\equal {#1}
        {}}{Ref.~\cite{#2}}{Ref.~\cite[#1]{#2}}}

\newcommand{\JF}{\texttt{JeLLyFysh}}
\newcommand{\Methods}{Methods\xspace}
\newcommand{\Supp}{Supplementary Material\xspace}
\newcommand{\ECMC}{event-chain Monte Carlo\xspace}
\newcommand{\LECMC}{Event-chain Monte Carlo\xspace}
\newcommand{\LJ}{Lennard-Jones\xspace}
\newcommand{\inclination}{\gamma}
\newcommand{\tmix}{t_{\text{mix}}}
\newcommand{\dropradius}{R}

\begin{document}

\title{Lifting the fog---a case for non-reversible \quot{lifted} Markov chains}

\author[a]{Gabriele Tartero}
\author[b]{Sora Shiratani}
\author[a,c,1]{Werner Krauth}

\affil[a]{Laboratoire de Physique de l'École normale supérieure, ENS,
    Université PSL, CNRS, Sorbonne Université, Université Paris Cité,
    Paris, France}
\affil[b]{Department of Physics, The University of Tokyo, Tokyo, Japan}
\affil[c]{Rudolf Peierls Centre for Theoretical Physics, Clarendon
    Laboratory, Oxford OX1 3PU, UK}

\leadauthor{Tartero}

\significancestatement{The Metropolis algorithm, as many mainstays of
computational science, is patterned after physical dynamics, and it thus
time-reversible. We discuss the limitations that arise from reversibility, and
how they are overcome through non-reversible \quot{liftings}. In the
\LJ liquid, where reversibility necessarily leads to slow Ostwald
ripening in the coarsening dynamics towards equilibrium, we discribe a lensing
effect that induces relative motion of macroscopic droplets in the lifted
algorithm, namely \ECMC. A highly efficient implementation for
long-range interactions allows us to demonstrate an increased coarsening growth
exponent and thus a reduced scaling for the mixing time. Therefore,
for large system sizes, the lifted algorithm converges infinitely faster than
the original reversible one.
Our findings explain recent successes of \ECMC
and point to future applications of non-reversible Markov chains in
computational science and applications.
}

\correspondingauthor{\textsuperscript{1}To whom correspondence should be
    addressed. E-mail: werner.krauth@ens.fr}

\begin{abstract}
Phase transitions appear all over science, and are familiar from everyday life,
as water boiling, sugar melting into caramel or as nematic molecules turning
smectic in liquid-crystal displays. The dynamics of phase transitions can be
extremely slow, as for example when fog in winter does not lift, that is when
the coarsening takes much time from many tiny water droplets to fewer but larger
rain drops that feel the pull of gravity. The dynamics of phase transitions is
relevant also for the performance of computer algorithms, whose time evolution
is often patterned after physical processes. In the ubiquitous Metropolis Monte
Carlo algorithm,
the mixing dynamics
towards equilibrium leads
towards the solution of a sampling problem. It is governed by the same
reversibility and detailed-balance principles as the overdamped physical
dynamics of fog. In the example of the phase-separated \LJ system, we
describe here how the coarsening dynamics of non-reversible \quot{lifted}
variants of the Metropolis algorithm proceeds on much faster time scales, with
the microscopic non-reversibility translating into large-scale relative motion
of droplets that is impossible under the Ostwald-ripening condition of
reversibility. A density--displacement coupling moves droplets relative to each
other through a lensing effect. Efficient implementations of the long-range
Metropolis algorithm and its non-reversible lifting (\ECMC)
allow us to show that, in consequence, the coarsening growth exponent is
larger under lifting. For large system sizes, the computing problem is thus
solved infinitely faster than before, with the outcome (the equilibrium state)
strictly
unchanged  with respect to the Metropolis algorithm. We also discuss the
larger setting of our findings, namely that \quot{lifted} non-reversible
algorithms can be set up for generic reversible sampling methods,
with applications going much beyond our example of lifting fog.
\end{abstract}

\dates{This manuscript was compiled on \today}
\doi{\url{www.pnas.org/cgi/doi/10.1073/pnas.XXXXXXXX}}

\maketitle
\thispagestyle{firststyle}
\ifthenelse{\boolean{shortarticle}}{\ifthenelse{\boolean{singlecolumn}}
{\abscontentformatted}{\abscontent}}{}

\firstpage[9]{2}

\begin{figure*}[tbhp]
    \centering
    \includegraphics[width=.8\linewidth]{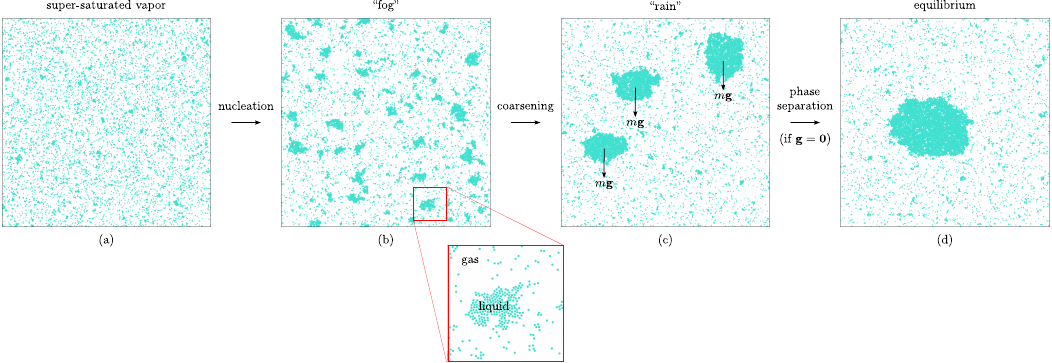}
    \caption{Coarsening of the two-dimensional \LJ system ($N=10^4$
particles at density $\rho = 0.1$) from super
saturation (a) into a single liquid drop surrounded by vapor
(d). After nucleating
several droplets (b), the system goes through a coarsening process,
that
gradually reduces the number of liquid clusters while increasing their average
size (c). In the absence of gravity, the equilibrium state contains a
single
floating drop. The Metropolis algorithm, that mimics the physics of
Ostwald ripening, needs about $10^{12}$ steps to
complete this process, while its lifted counterpart, \ECMC,
reaches equilibrium after around $10^9$ events.}
     \label{fig:Coarsening}
\end{figure*}

\dropcap{F}or a range of temperatures and volumes, the equilibrium state of a
system of $N$ particles may be composed of two coexisting phases, following
their separation at a first-order transition. During a cool night, for example,
part of the water vapor contained in  air may nucleate into tiny liquid
droplets, creating fog. In spite of its persistence in time, familiar to every
reader of 19th century English novels, fog is far from being an equilibrium
state: It is unstable at large times. The explanation for this inescapable fact
is that Nature strives to minimize free energy and that a single drop has
smaller surface and lower surface free energy than two droplets of same combined
volume.
Eventually, droplets must thus coalesce until (in a finite volume) they form a
single approximately spherical drop or until they precipitate as rain under the
action of gravity (see \fig{fig:Coarsening}). The persistent and apparently
paradoxical Dickensian fog of \emph{Bleak House}~\cite{Dickens2008}, sticking
around for days, illustrates that the coarsening dynamics of droplets increasing
in size and decreasing in number is excruciatingly slow. Fog practically never
precipitates into rain, and it generally lifts through a change of the external
parameters: with an increase in temperature or with wind that starts to blow.
After a while, a new equilibrium state is attained at a different temperature,
without the original phase separation.

Computational science patterns many of its techniques after physical processes.
Thus, molecular dynamics and gradient descent mimic Newtonian time evolution in
applications from physics to machine learning, and the vast field of Monte Carlo
sampling, pioneered in physics by Metropolis et al. in 1953, now permeates all
of science. The Metropolis algorithm, one of the most cited works in
science~\cite{Metropolis1953}, is a black box for computing and is perceived as
a panacea for arbitrary sampling problems. The Metropolis algorithm implements
time-reversible dynamics that converges towards equilibrium. The time
reversibility, in other words the detailed-balance condition, built into the
algorithm in or out of physics, is identical to the governing principle of the
fog dynamics of tiny droplets. In very general applications, the Metropolis
algorithm will, so to speak, eventually turn small droplets into rain, but
the mixing time~\cite{Levin2008} for this to come to fruition can be enormous,
just as for the
\quot{fog everywhere} in \quot{implacable November
weather}~\cite[p.12]{Dickens2008}. Two aspects of the analogy with London fog
are crucial: First, the equilibrium state encoded in the Metropolis algorithm
represents the solution of a given computational problem, and we require a
speed-up of the dynamics without changing the steady state (without having the
sun dissipate the fog). Second, in the case of a two-phase mixture, it has been
understood since Ostwald in the late 1800s that a configuration
with several droplets cannot be stationary, thus providing a clear-cut
criterion for stationarity. In general, however, it is very difficult to
ascertain
that the steady state has been reached. This risks flawed conclusions,
as for example  in \subfig{fig:Coarsening}{b}
(if we lacked Ostwald's insight), which is \quot{almost} stationary
under local reversible dynamics, but which is not at all in equilibrium
(represented by \subfig{fig:Coarsening}{d}).

Non-reversible (thus, effectively, non-equilibrium) Markov chains violate
detailed balance, and they feature finite steady-state flows even
if the steady state is the equilibrium Boltzmann distribution. The perceived
paradox of non-equilibrium in equilibrium  has  been clarified in exactly
solved models ~\cite{KapferKrauth2017,Essler2024}. The practical construction
of non-reversible Markov chains with an imposed steady state has long lacked
clear design principles, but the concept of \quot{lifting}, which copies every
configuration of a parent chain into multiple configurations, has allowed
definite progress~\cite{Diaconis2000,Chen1999}. An example of non-reversible
lifted Markov chains is given by the \ECMC algorithms
for particle systems whose Boltzmann distributions then
appears as a stationary state with finite
flows~\cite{Krauth2021eventchain,Peters2026}.

The characteristic convergence times for lifted Markov chains obey mathematical
constraints~\cite{Chen1999}. First, in order to speed up a reversible parent
Markov chain,
its lifting must be non-reversible. This validates our intuition that the slow
evaporation of liquid particles from a small droplet into the vapor and the
condensation of vapor particle to a larger liquid droplet
(Ostwald ripening~\cite{Lifshitz1961, Wagner1961}) explains
the persistence of fog, and more generally the slowness of the local Metropolis
algorithm. Second, the convergence times of a lifted Markov chain are lower
bounded by the square root of the times of its reversible parent algorithm. This
validates the concept of ballistic \vs\ diffusive motion for general Markov
chains, much beyond the realm of particle systems. However, the lower bound does
not guarantee that non-reversibility at a microscopic level actually speeds up
convergence. Up to now, it has remained unclear whether a
speed-up beyond a size-independent factor can be obtained in real-life
many-particle systems beyond one dimension.

In this paper, we first discuss reversible and non-reversible Markov chains,
with, in the case of long-range-interacting particle systems, the (factorized)
Metropolis algorithm as an example of the former and \ECMC of the latter.
Second, we analyze how
\ECMC accelerates the coarsening dynamics of a two-dimensional \LJ system
through fast
relative motion of liquid droplets, what the Metropolis algorithm cannot
achieve.
The limitations of Ostwald ripening are
overcome by a density--velocity
coupling that induces macroscopic motion of droplets with respect to each other.
A related coupling has been analyzed rigorously in a one-dimensional
context~\cite{Massoulie2025}. Third, we provide highly optimized open-source
implementations of these algorithms, to rigorously simulate cutoff-free
\LJ particle systems. This allows us to quantify the above effects and
to show that, on a macroscopic scale, they lead to a better coarsening
growth exponent
for this process (characterizing the mean droplet size as a function of time).
We relate the coarsening growth exponent to the much improved scaling of the
mixing time with system size. In the conclusions,
we review from a larger perspective why non-reversibility is
responsible for the improved scaling and the
infinite speedup of the non-reversible algorithm for large systems.
We discuss possible applications of our findings in statistical physics, and
extensions to chemical physics and to machine learning.

\section*{Reversible and non-reversible coarsening}

In the fog of 19th century London, small droplets move in over-damped (windless)
conditions and the Maxwell velocity of macroscopic droplets effectively
vanishes, together with all probability flows. The detailed-balance dynamics is
time-reversible,
just as the Metropolis algorithm. To showcase reversible and non-reversible
coarsening in a model that is easily visualized, we pit against each
other the factorized Metropolis algorithm~\cite{Michel2014JCP} (which
is reversible) and the \ECMC algorithm (which is lifted and
non-reversible),  in the two-dimensional \LJ model with
long-range interactions (see \Methods for details on the algorithm and the model
parameters). At each time step, the Metropolis algorithm attempts to move a
random particle symmetrically around its current position. The lifted algorithm,
\emph{mutatis mutandis} makes the same moves but organizes them
differently. For a certain time, they are all in the $+ x$
direction, then in
the $+ y$ direction (the moves in $-x$ and $-y$ are not required in a periodic
system). After an accepted move of particle $i$ it moves that
particle again. Because of the factorization, rejection can always be
attributed to a single target particle $j$. After a
move of active particle $i$ rejected by target particle $j$, it is $j$ that
moves next, and in the same
direction. After a certain time, the direction of motion switches between $+x$
and $+y$. An additional \quot{factor-field} control
parameter~\cite{Lei2019, Maggs2024} of \ECMC will not be explored in this
paper.

\LECMC converges to a NESS (\quot{non-equilibrium steady
state}) identical to  equilibrium, yet it conserves finite particle and
probability flows. Two of our
movies illustrate the time evolution from the super-saturated vapor towards
phase-separated equilibrium configurations. For the reversible algorithm,
macroscopic droplets are essentially immobile, but larger droplets
generally grow at the
expense of smaller ones that shrink, then disappear, thus realizing
Ostwald ripening
(see Movie S1 in the \Supp). This is because smaller droplets have a slightly
higher
particle evaporation rate yet a lower particle adsorption rate than bigger ones.
The non-reversible algorithm has
droplets move
with respect to each other. This allows them to coalesce (see Movie
S2). In a system of $N=10^4$ \LJ particles at density
$\rho=\fraca{N}{L^2}=0.1$ (corresponding to \fig{fig:Coarsening} and
Movies S1, S2 and S3),
the coarsening process ends with a single droplet of size $\sim 0.085 \times
L^2$ (that is, a radius of $R \simeq 52.0$ in a box of length $L \simeq 316.2$)
after $\sim 10^{12}$
individual steps for the reversible Metropolis algorithm and after $\sim 10^9$
steps for non-reversible \ECMC. The steady states of
these algorithms are rigorously the same. The speedup is substantial.

\section*{Density--velocity coupling of non-reversible dynamics}

\begin{figure*}[tbhp]
\centering
\includegraphics[width=.8\linewidth]{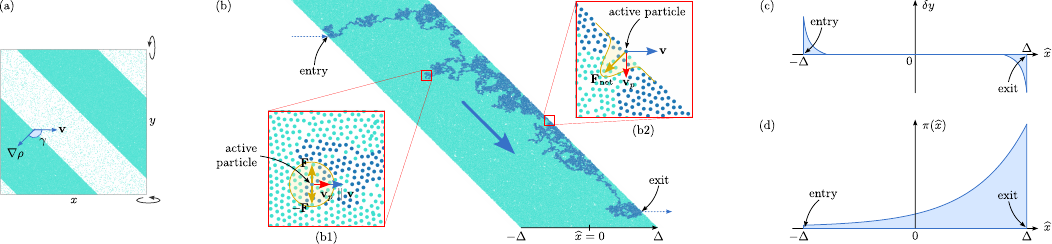}
\caption{Density--velocity coupling of \ECMC in a periodic
liquid ribbon.
\subcap{a} The ribbon in its periodic simulation box ($\sim 10^5$
liquid particles at density $\rho=0.75$ surrounded by vapor, see \Supp for
details). The event chain's
direction of motion is towards the right ($+x$)
\subcap{b} Typical event chain, from its entry into the ribbon to its exit from
it.
\subcap{b1} Inside the ribbon, the active particle interacts---and the
event
chain grows---symmetrically around the direction of motion. \subcap{b2} At the
ribbon boundary, the active particle interacts symmetrically around the density
gradient and the event chain grows with a component perpendicular to the
direction of motion (here, in the $-y $ direction).
\subcap{c} Mean vertical growth component $\delta y$ of the event chain as a
function of its horizontal position $\xhat \in [-\Delta, \Delta]$ in the
ribbon. The opposite boundary effects are due to the opposite sign of the
density gradient at the two interfaces.
\subcap{d} Probability distribution of the active particle's $\xhat$ coordinate.
The event chain grows predominantly near the forward boundary, as already
evident from (b). The curves in panels (c) and (d) are smoothed, schematic,
versions of the plots from the \Supp.
}\label{fig:Ribbon}
\end{figure*}

To analyze the speedup of the coarsening dynamics (Movie~2 compared to Movie~1),
we consider the growth of an event chain at a density gradient with inclination
angle $\inclination$ with respect to its direction of motion. For concreteness,
we realize the density gradient
as an infinite
periodic ribbon of liquid surrounded by vapor and take
the
direction of motion as $+x$ (see \subfig{fig:Ribbon}{a}). The active particle
$i$ at the head of the event chain moves
in $+x$ until its further displacement is opposed by the \quot{target} particle
$j$ which in turn becomes active. The target particle is most likely inside the
ribbon and in the direction of the density gradient, simply because that is
where most candidate target particles are situated. If the ribbon is oriented
as in \subfig{fig:Ribbon}{a} (that is, for $ \pi/2 <\inclination  <
\pi$),
the target particle and, thus, the new head of the event chain will likely be
at a lower $y$. After the entry of the event
chain into the ribbon, the chain first grows along the direction
of motion, but close to
the right boundary, it repeatedly grows in direction of the density
gradient, that is, downward (see \subfig{fig:Ribbon}{b}). Two aspects of the
effect stand out. First, at the back and the front of the ribbon,
the event chain is deflected perpendicularly to the direction of motion
(\subfig{fig:Ribbon}{c}). This creates a macroscopic perpendicular offset of
the
exit position with respect to the entry point. Second, particles at the
front of the ribbon are more likely to participate in the growth of the event
chain than particles in its center or at its back (see \subfig{fig:Ribbon}{d}).
This will end up by destroying the ribbon because particles in the event chain
move forward whereas the others do not. It also indicates that the ribbon must
be out of equilibrium as, in equilibrium, each particle in the
system has the same probability to be active (see \Methods).

\section*{Lensing}

\begin{figure*}[tbhp]
\centering
\includegraphics[width=.8\linewidth]{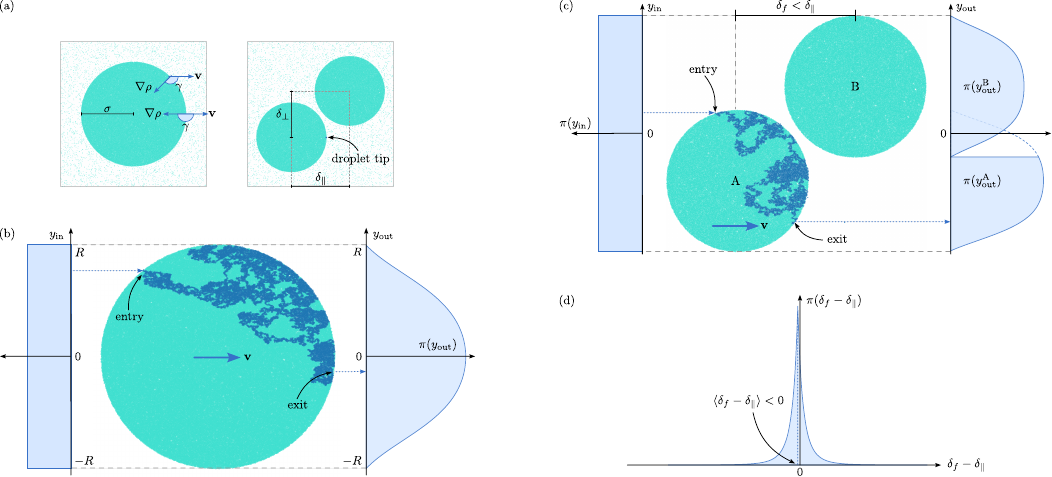}
\caption{Lensing effect of \ECMC and relative motion of
droplets.
\subcap{a} Circular liquid droplet of radius $\dropradius$ and pair of
droplets in their periodic
simulation
boxes.
\subcap{b} Lensing effect of event chain Monte Carlo inside a liquid drop. The
event chain,
entering the drop uniformly undergoes multiple gradient effects and then
leaves it predominantly near its tip.
\subcap{c} Relative motion of two droplets created by the lensing effect. An
event
chain entering drop A, even at its extremity, likely exits it without touching
drop B (see histogram). The expected time spent in A, and thus the displacement
of A in $+x$ direction is larger than the time spent in and the displacement of
B.
\subcap{d} Probability
distribution of the variation $\delta_f - \delta_\parallel$ in the horizontal
separation between the droplets. Its negative mean value shows that the two
droplied approach each other. The curves in panels (b), (c) and (d) are
smoothed versions of histograms from the \Supp.
}\label{fig:LensingEffect}
\end{figure*}

In the ribbon of \fig{fig:Ribbon}, the density gradient deflects the event chain
perpendicularly to the direction of motion in a way that is strongest when the
ribbon is almost parallel to the direction of motion (for an inclination angle
$\inclination \gtrsim \pi/2$), and that vanishes when it is almost perpendicular
(inclination angle $\inclination \sim \pi$). Inside a spherical liquid droplet,
the event chain encounters a varying inclination angle, possibly starting with
strong deflection (near the north or south poles in
\subfig{fig:LensingEffect}{b}), then ending close to the tip
of the droplet with $\inclination \sim \pi$ and no deflection (see
\subfig{fig:LensingEffect}{b}). This results in a migration of
the event chain from the poles towards the tip of the droplet, in other words
in a lensing. In
our example of the motion in $+x$, a uniform distribution of  event-chain entry
heights in $y$ produces a distribution of exit heights $y$ that is centered
around the symmetry axis (see \subfig{fig:LensingEffect}{b}). The lensing effect
exists for general potentials and even for hard disks. We expect it
to depend on the factor-field control parameter of \ECMC.

\section*{Relative motion of droplets}

\begin{figure*}[tbhp]
\centering
\includegraphics[width=.8\linewidth]{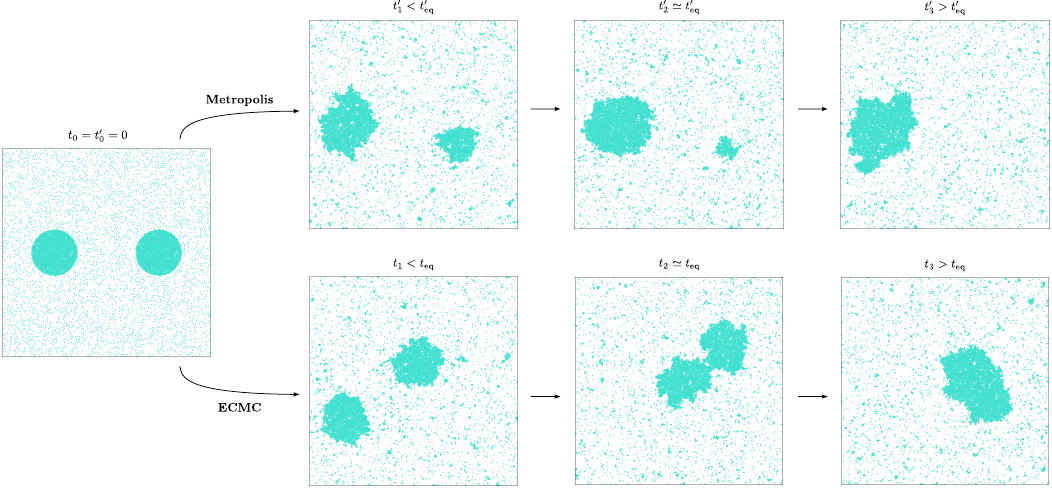}
\caption{Equilibration dynamics of the factorized Metropolis algorithm and of
\ECMC from an initial configuration with two spherical
droplets (with 2565 particles each, at local density $\rho=0.75$)
surrounded by 3872  vapor particles at local density $\rho=0.05$, and
overall density $\rho=0.1$. Under the Metropolis
dynamics, the system equilibrates through Ostwald ripening, that is, through the
evaporation--condensation of vapor at the surface of immobile liquid
droplets (here, after around $10^{11}$ moves). Under \ECMC dynamics,
the two droplets move relative to each other, allowing them to coalesce
much faster (here, after about $10^8$ events). We attribute this acceleration
to the lensing effect.
}\label{fig:MetropolisVsECMC}
\end{figure*}

Because of the lensing, event chains that enter a droplet A of radius
$\dropradius$ migrate towards its tip and then exit. The time they spend in
droplet A determines how much the latter advances. We consider a second droplet
B (also of radius $\dropradius$) in the shadow of A,
for a perpendicular
pair distance $\delta_\perp$ smaller than twice the droplet radius
($\delta_\perp < 2 \dropradius$)
(see
\subfig{fig:LensingEffect}{a}). The droplet A then  advances for a wide
choice of entry
positions, and the event chain exits without touching B, although it is in its
shadow. If any event chain
was exiting A exactly at its tip (complete lensing),
droplets A and B would on average approach each
other for $2 \dropradius > \delta_\perp > \dropradius$ and would move away from
each
other for $ \dropradius > \delta_\perp > 0$. This is because any chain entering
droplet A (and advancing it along the direction of motion)
would then also enter droplet B. In practice,
the effect of the shadowing is obtained by numerical simulation (see
\subfig{fig:LensingEffect}{b})
and different perpendicular distances cause the droplets to more or
less move with respect to each other (see
\fig{fig:LensingEffect}{(d)} for a histogram and \Supp for a discussion of the
different possible cases). Lensing
thus overcomes the immobility constraint of Ostwald ripening. This
effect survives on the longest time scales as can be tested by relaxing a
configuration with two equal-sized droplets all the way into equilibrium and by
comparing typical evolutions for the factorized Metropolis algorithm and for
\ECMC (see \fig{fig:MetropolisVsECMC}).

\section*{Coarsening growth exponents}
\begin{figure}[tbhp]
\centering
\includegraphics[width=1.0\linewidth]{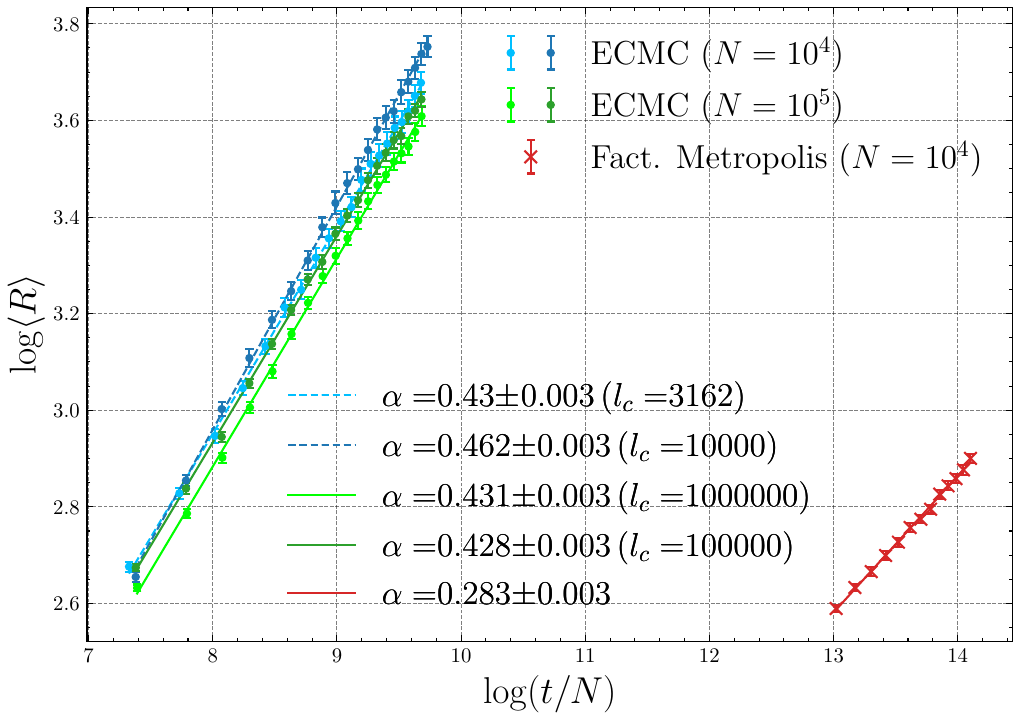}
\caption{Coarsening dynamics of \ECMC and factorized Metropolis for a
two-dimensional \LJ system in the
phase-coexistence regime ($\rho = 0.1$). The average droplet radius
$\mean{R}$ depends on time as a power law of growth exponent $\alpha$,
with the time $t$ corresponding to the total
distance traveled by all particles in the system.
\LECMC outperforms Metropolis both in terms of scaling and of
absolute CPU time.
Yet, for finite systems, the growth exponent $\alpha$ is found to
depend on the event-chain length.
Roughly, \ECMC mixes 100 times faster than the factorized Metropolis
algorithm for $N=10^4$, and 1000 times faster for $N=10^5$.
}
\label{fig:logRlogt}
\end{figure}

The lensing effect of  \ECMC has consequences at the largest
length and time scales, as evidenced in its coarsening dynamics,
in other words in the growth with time of the mean radius $\mean{R}$ of liquid
droplets surrounded by vapor. During coarsening (after the nucleation of
many tiny droplets from the super-saturated vapor), the evolution of the mean
radius generally follows a power
law $\mean{R} \propto t^\alpha$, where $\alpha > 0$ is the growth exponent. When
coarsening proceeds by evaporation of single particles from smaller
immobile droplets and
condensation of single particles onto larger immobile ones (Ostwald ripening),
the coarsening growth exponent is
$\alpha = \fraca{1}{3}$ both in two and three
dimensions~\cite{Lifshitz1961, Wagner1961, Zheng1989, Das2015}.
We indeed observe a very small growth exponent close to the theoretical value
in our \LJ model  for the Metropolis
algorithm (see \fig{fig:logRlogt}). The lack of mobility of droplets
is readily evidenced in Movie S1. Similar behavior was also
numerically verified for the two-dimensional Ising
model under Metropolis dynamics~\cite{Majumder2010}.

The \ECMC dynamics allows for the
macroscopic motion of the droplets themselves. This leads to a larger growth
exponent, whose value is influenced by the strength of the lensing effect. For
well-chosen values of the chain length $l_c$, the growth exponent increases
significantly, thus making \ECMC infinitely faster than
Metropolis in the thermodynamic limit (see \fig{fig:logRlogt}).
We may identify the time to arrive at a single
droplet of size $\propto L$ with the mixing time of the Markov chain: $\mean{R}
\sim
(t/N)^\alpha$.
For $R \sim L$, thus $\mean{R}^{1/\alpha} \sim t/L^2$, we then reach,
\begin{equation}
\tmix  \sim  L^{2 + 1/\alpha}.
\end{equation}
Here, the term  in $L^2$ (which is specific to two spatial dimensions) reflects
that we count time in single displacements rather than in sweeps. A larger
coarsening growth exponent then implies an improved scaling of the mixing time.
The empirical growth exponent varies with $\lch$, likely because the latter
influences the lensing. Too small $\lch$ do not allow the chain to grow and a
sufficient number  of particles to be displaced inside a droplet. On the other
hand, too large $l_c$ would hinder the relative droplet displacement due to
periodic boundary conditions. In the present paper, we only consider
implementations where $\lch$ is fixed throughout the simulation. A dynamically
changing $\lch(t) \sim R(t)^2 $ (which visits one droplet per chain) may lead
to smaller mixing times than a fixed chain length $\lch \sim N$, which might
furthermore be influenced by factor fields. Nucleation preceeds coarsening on a
time scale (in our units) of $L^2$, which is asymptotically negligible compared
to the coarsening time scale.

\section*{Non-reversible steady-state dynamics}
The out-of-equlibrium dynamics of \ECMC, as we have
demonstrated for specific initial configurations as the ribbon and the
spherical droplet, leads to an
inhomogeneous growth of event chains, and to an inhomogeneous
probability distribution of active particles. It is this inhomogeneous
distribution which causes the lensing and which allows the algorithm
to overcome the limitations of Ostwald ripening.
In equilibrium, the probability
distribution of the active particles over the system is uniform, by
construction principle for
lifted Markov chains (see \Methods for a discussion). In consequence, the
motion of the single remaining liquid comes to a still-stand, and it
hardly moves with respect to the surrounding vapor (see Movie S3).
This fascinating change of behavior between out-of-equilibrium and
equilibrium dynamics links the equilibrium geometry of the liquid--vapor
interface to the behavior of \ECMC in a way that is not yet understood.

\section*{Conclusion and outlook}
In this paper, we have studied the transmission of non-reversibility in local
Markov chains from the smallest to the largest length and time scales. In our
example of the two-dimensional \LJ system approaching coexistence, we
showed how non-equilibrium density fluctuations influence the mixing dynamics.
The relative motion of droplets is a consequence of an intrinsic
density--velocity
coupling, that arises naturally in lifted algorithms, and that cannot be
achieved in reversible Monte Carlo algorithms and neither in molecular dynamics
nor in Hamiltonian Monte Carlo~\cite{Krauth2024hamiltonian}.
The resulting dynamics mixes much faster than the reversible Metropolis
algorithm, where droplets barely move and are thus doomed either to evaporate
or to grow, until equilibrium is reached.
Quantitatively, we were able to show that, during coarsening,
\ECMC exhibits a larger growth exponent
than Metropolis. As a consequence,
its mixing time has better scaling than the reversible one.

Analogous behavior was found in a one-dimensional particle system on a
lattice under what amounts to \ECMC dynamics~\cite{Massoulie2025}. The coupling
of density differences to the local velocity there speeds up the
mixing dynamics for specific initial configurations as well as the relaxation
dynamics for equilibrium density fluctuations. The extension to two spatial
dimensions that we presented in this paper is remarkable, and it would be
interesting to understand in more general terms the role of the
dimension. The lensing effect may well explain the effectiveness of \ECMC
algorithms in hard-disk systems~\cite{Bernard2011} which are governed by
the physics of coarsening of hexatic droplets in a liquid background rather
than that of liquid droplets in a vapor background studied in the
present paper.
In essence,  we have shown here that a non-reversible \emph{local}
Markov chain can perform better ($100$ times for $N= 10,000$, $1000$ times
for $N= 100,000$) than a reversible \emph{local} Markov chain, in the real-life
setting of the \LJ system. What is gained cannot be achieved with simply
replacing reversible Markov chains with molecular dynamics, Hamiltonian Monte
Carlo or stochastic gradient descent, as was argued
elsewhere~\cite{Krauth2024hamiltonian}. We expect applications of
non-reversible lifted Markov chains from chemical
physics to the vast field of machine learning.

\matmethods{

\subsection*{\LJ system}
In this paper, we consider a system of $N$ point particles in a periodic
two-dimensional square box of size $L$, interacting via the \LJ pair potential:
\begin{equation}
\Ulj(r) = 4 \varepsilon \glc \glb \fracb{\sigma}{r} \grb^{12} -
\glb \fracb{\sigma}{r} \grb^6 \grc.
\label{equ:LennardJones}
\end{equation}
The equilibrium Boltzmann weight of a configuration is  $\expc{-\beta
\sum_{i<j=1}^N \Ulj(r_{ij})}$, where $\beta$ is the inverse temperature, and
$r_{ij}$ is the periodically corrected pair distance between particles $i$ and
$j$. We introduce no cutoff in the model, but nevertheless arrive at efficient
reversible and non-reversible implementations (see below for more details). In
\eq{equ:LennardJones}, we set $\sigma = 1$, $\varepsilon=\fraca{1}{0.46}$ for
every system size, and consider a constant number density $\rho = N/ L^2 = 0.1$
for all system sizes.  At inverse temperature $\beta = 1$, these values lead to
a phase-separated equilibrium state, where the liquid particles are gathered in
a single drop of local density $\rho \simeq 0.71$ surrounded by vapor of
local density $\rho \simeq 0.04$ \cite{LiCiamarra2020b}. To identify the
liquid during the time evolution, we coarse-grain the system with a regular grid
with cells of volume 100 and classify as \quot{liquid} any cell whose local
density is larger than $\rho=0.5$ (that is, which contains more than 50
particles). A liquid droplet then constitutes a contiguous set of liquid cells
of a well-defined coarse-grained two-dimensional area, the square root of
which, divided by $\sqrt{\pi}$, we take to be the radius $R$. The mean radius
$\mean{R(t)}$ of \fig{fig:logRlogt} takes an average of $R$ over the droplets in
a given configuration.

\subsection*{Lifted Markov chains}

Lifting~\cite{Chen1999} connects an original reversible Markov chain, in our
case the factorized Metropolis algorithm, with a Markov chain in an augmented
space in a way that reorganizes the moves. If the Metropolis algorithm proposes
a random (forward or backward, upward or downward) move of a random particle,
its lifted version keeps memory of the previous move (say, forward) and of the
previous particle $i$. The original sample space is thus augmented by the
direction and by the index of what we refer to in the main text as the
\quot{active} particle.

A lifted Markov chain requires the statistical weight of all the \quot{lifted}
copies $\ahat$ of an original configuration $a$ (that is, the configuration $a$
of particles together with the direction and the index of the active particle)
to add up to the statistical weight (the Boltzmann weight) of $a$. More
restrictively, we require all lifted copies $\ahat$ of $a$ to have the same
equilibrium stationary weight. This condition makes that if, in
\subfig{fig:Ribbon}{b}, particles at the boundary of the ribbon are more likely
to be active, this ribbon configuration cannot be in equilibrium.
In addition to the equal-repartition rule, one requires also
that the statistical flows (the stationary weights $\pi(\ahat)$  multiplied with
the transition probabilities  $P(\ahat \to \bhat)$) sum up to the statistical
flows between the original configurations $a$ and $b$. Under these conditions of
lifting, it has become possible to systematically construct non-reversible
Markov chains. Lifting can reduce mixing and relaxation times of an original
Markov chain roughly to their square roots~\cite{Chen1999,Krauth2021eventchain},
but the lifted Markov chain must then be non-reversible.

\subsection*{Metropolis algorithm with long-range interactions}

The traditional Metropolis algorithm proposes moves, in our case displacements
of single particles between configurations  $a $ and $b$, such
that the move $a\to b$ is proposed with the same probability as the move $b \to
a$. Moves are accepted on the basis of the relative Boltzmann weights of $a$ and
$b$, that is, in terms of the change of the total energy difference $\Delta U$
between $b$ and $a$. For the long-range pair potential $\Ulj$ of
\eq{equ:LennardJones} $\Delta \Ulj$, for a single-particle displacement, has
contributions from throughout the system, resulting in a naive complexity
\bigOb{N}
per single-particle move. As in molecular dynamics, this problem is commonly
overcome by cutting off the potential at a finite spatial range, so that the
displacement of a single particle impacts only a finite number of nearby pairs
of particles. An \bigOb{1} complexity per move is obtained, but the
required extrapolations to infinite range are often
fraught with uncertainty~\cite{Smit1991, Smit1992, Lei2005}.
Here, we rather use the factorized Metropolis algorithm~\cite{Michel2014JCP},
which accepts or rejects moves on the basis of a consensus of Metropolis
decisions for all pair factors impacted by the displacement of a single
particle. The method can be implemented in \bigOb{1} per move without
cutoff~\cite{Tartero2024JCP}. For all intents and purposes, it agrees with
the traditional Metropolis algorithm (see also
\REF{Tartero2024AJP}). A fast \bigOb{1} implementation derived from an
open-source Python code is used in this paper (see below for access to the
code).

\subsection*{\LECMC with long-range interactions}
Event-chain
Monte Carlo represents a lifting of the factorized Metropolis algorithm in the
limit of infinitesimal moves~\cite{Krauth2021eventchain}. As described in the
main text,
a sequence of repeated infinitesimal Metropolis moves of a given \quot{active}
particle
are proposed in a fixed direction, until one such move is rejected. Any
rejection is
attributed to a definite \quot{target} particle, which then becomes the active
one. With periodic boundary conditions, it is only necessary to have
moves in the $+x$ and $+y$ direction.
To ensure irreducibility, after a certain chain length $l_c$,
a new active particle is randomly sampled and the direction is switched
between $+x$ and $+y$. With infinitesimal moves, the algorithm becomes
event-driven, and it generates the eponymous event chains. For hard spheres,
this
amounts to computing the position of the next collision, given the direction of
motion~\cite{Bernard2009}. For continuous interactions,
the lifted factorized Metropolis algorithm allows the target particle to be
sampled exactly~\cite{Peters_2012,Michel2014JCP}.
Remarkably, this allows the new active particle to be sampled  with a
complexity
\bigOb{1}~\cite{KapferKrauth2016,Tartero2024}. The Boltzmann distribution is
sampled without any correction (for an introduction to this method, see
\REF{Tartero2024AJP}).

\subsection*{Software packages}
The numerical simulations for this paper (factorized Metropolis algorithm and
\ECMC for the \LJ system without cutoff)
were performed with open-source code. Our
implementations
are based on the programs provided with \REF{Tartero2024JCP}. They
all have \bigOb{1} per move complexity.
Our codes were carefully checked
against \JF, an open-source Python application that implements \ECMC
for general short- and long-range interactions~\cite{Hoellmer2020}.
Our \ECMC program written in the Rust language is available at
\url{https://github.com/EarlMilktea/lj-ecmc.git}.

\subsection*{Supporting information}
Our \ECMC simulations of \fig{fig:LensingEffect} are
complemented  in the \Supp  by a number of plots for the lensing effect for a
single droplet and for the relative motion of pairs of droplets under varying
shadowing conditions. In addition, the \Supp also contains three movies that
illustrate
the different reversible and non-reversible Monte Carlo dynamics for the
two-dimensional \LJ system. They are as follows:
\begin{description}
 \item[Movie S1] (\quot{Coarsening dynamics of the factorized Metropolis
algorithm})  starts from an
all-vapor initial configuration with $N= 10,000$. It then undergoes nucleation
and coarsening to arrive at the equilibrium, as in \fig{fig:Coarsening}. The
total time to arrive at the equilibrium state with a single droplet is
$\sim 10^{12}$ moves, that is, $\sim 10^8$ moves per particle. The Ostwald
ripening is
manifest, as the big droplets hardly move. They grow through the
evaporation of single particles from the small droplets into the vapor and the
condensation of single vapor particles onto larger droplets.
\item[Movie S2] (\quot{Coarsening dynamics of non-reversible \ECMC}) also
starts
from an all-vapor initial configuration. Equilibrium is reached
through nucleation and coarsening on a much shorter timescale
(around $10^5$ events per particle). The relative droplet motion manifestly
overcomes the limitations of Ostwald ripening.
\item[Movie S3] (\quot{Equilibrium dynamics of non-reversible \ECMC}) starts
from
an equilibrated initial configuration. The equilibrium drop hardly moves, in
keeping with the basic property of lifted Markov chains which, in equilibrium,
displace each particle with the same probability.
The difference between the equilibrium (this movie) and
non-equilibrium dynamics (Movie S2) is striking.
\end{description}

All movies track particle configurations in a center-of-mass reference frame,
where the mean displacement vanishes for all times.
}

\showmatmethods{} 

\acknow{
W.K. acknowledges generous support by the Leverhulme Trust.
Part of this research was conducted within the context of the International
Research Project \quot{Non-Reversible Markov chains, Implementations
and Applications}. S.S. thanks the Laboratoire de Physique de l'École normale
supérieure for hospitality. G.T. is grateful to Synge Todo and Koji Hukushima
for helpful discussions and insights. We thank Kush Patel for help with
the efficient implementation of the factorized Metropolis algorithm.
}

\showacknow{} 

\bibsplit[15]


\end{document}